\pdfoutput=1
\documentclass[11pt]{article}
\usepackage{acl}%commit
\usepackage{multirow}
\usepackage{enumitem}
\usepackage{array}
\usepackage{latexsym}
\usepackage{graphicx}
\usepackage{subfigure}
\usepackage{enumitem}
\setlist[itemize]{itemsep=0pt, parsep=0pt, leftmargin = 0pt}
\usepackage{amsthm,amsmath,amssymb}
\usepackage{mathrsfs}
\usepackage{booktabs}
\usepackage{float}%提供float浮动环境
\usepackage{booktabs}%提供命令\toprule、\midrule、\bottomrule
\usepackage{multirow}%提供跨列命令\multicolumn{}{}{}
\usepackage{graphicx}
\usepackage{times}
\usepackage[T1]{fontenc}
\usepackage[utf8]{inputenc}

\usepackage{microtype}
\usepackage{inconsolata}
\title{HyCoRec: Hypergraph-Enhanced Multi-Preference Learning for Alleviating Matthew Effect in Conversational Recommendation}
\author{{Yongsen Zheng}\textsuperscript{\textnormal{1}}\textbf{,} {Ruilin Xu}\textsuperscript{\textnormal{1}}\textbf{,} {Ziliang Chen}\textsuperscript{\textnormal{2,5}}\textbf{,} {\textbf{Guohua Wang}}\textsuperscript{\textnormal{3}*}\textbf{,}\\
{\textbf{Mingjie Qian}}\textsuperscript{\textnormal{1}}\textbf{,} {\textbf{Jinghui Qin}}\textsuperscript{\textnormal{4}}\textbf{,} {\textbf{{Liang Lin}}\textsuperscript{\textnormal{1,2}\textbf{*}}}\\
\textsuperscript{\textnormal{1}}Sun Yat-sen University, 
\textsuperscript{\textnormal{2}}Peng Cheng Laboratory,
\textsuperscript{\textnormal{3}}South China Agricultural University\\ 
\textsuperscript{\textnormal{4}}Guangdong University of Technology,
\textsuperscript{\textnormal{5}}Jinan University\\
 \{z.yongsensmile, wangguohuagmial, scape1989\}@gmail.com\\
 \{xurlin5, qianmj7\}@mail2.sysu.edu.cn, c.ziliang@yahoo.com, linliang@ieee.org \\
}

\begin{document}
\maketitle
\renewcommand{\thefootnote}{}
\footnotetext{\textsuperscript{\textnormal{*}}Corresponding author. \\ 
}

\begin{abstract}
The Matthew effect is a notorious issue in Recommender Systems (RSs), \emph{i.e.}, the rich get richer and the poor get poorer, wherein popular items are overexposed while less popular ones are regularly ignored. Most methods examine Matthew effect in static or nearly-static recommendation scenarios. However, the Matthew effect will be increasingly amplified when the user interacts with the system over time. To address these issues, we propose a novel paradigm, Hypergraph-Enhanced Multi-Preference Learning for Alleviating Matthew Effect in Conversational Recommendation (HyCoRec), which aims to alleviate the Matthew effect in conversational recommendation. Concretely, HyCoRec devotes to alleviate the Matthew effect by learning multi-aspect preferences, \emph{i.e.}, item-, entity-, word-, review-, and knowledge-aspect preferences, to effectively generate responses in the conversational task and accurately predict items in the recommendation task when the user chats with the system over time. Extensive experiments conducted on two benchmarks validate that HyCoRec achieves new state-of-the-art performance and the superior of alleviating Matthew effect. Our code is available at {\color{blue}https://github.com/zysensmile/HyCoRec}.
\end{abstract}

\section{Introduction}
Conversational Recommender Systems (CRSs) engage in iterative conversations with users to provide personalized recommendations \cite{dialogue_1,dialogue_2,dialogue_3}, which have been widely adopted in various domains such as music recommendation \cite{music_conversation} and online e-commerce \cite{e_commerce}. Nevertheless, CRSs often face the prominent issue of Matthew effect \cite{MatthewEffect1}, which can be described as ``the rich get richer and the poor get poorer''. This phenomenon indicates that popular items/categories from past data receive more visibility in subsequent recommendations while less popular ones tend to be overlooked or ignored.

Recently, many research efforts have focused on examining the Matthew effect in static or relatively-static offline recommendation scenarios \cite{MatthewEffect1,MatthewEffect1_1,MatthewEffect1_2}. These offline studies strive to explore the potential causes behind the manifestation of the Matthew effect, and two key causes have been identified. One cause \cite{MatthewEffect1_1,MatthewEffect1_2,MatthewEffect1_3,MatthewEffect1_4} is that individuals with narrower and less diverse preferences exhibit a higher vulnerability to being trapped within the confines of the Matthew effect. Another cause \cite{MatthewEffect2_1} is that the severe popularity bias where popular items consistently receive amplified exposure while less popular ones are underexposed. Although these methods have undoubtedly contributed valuable insights into the phenomenon of the Matthew effect, they directly overlook the adverse impact stemming from the dynamic user-system feedback loop. More recently, Gao et al. \cite{MatthewEffect2} explore the Matthew effect in dynamic user-system interactions, but it lacks real-time user engagement through natural language conversations.

Despite their effectiveness, most methods still suffer from two major limitations. \emph{1) Interactive Schema}. Many methods aim to mitigate Matthew effect in the static recommendation settings without considering the user-system feedback loop \cite{feedback_loop}. In reality, the Matthew effect will progressively amplify as users dynamically interact with the system over time. Worse still, such amplification will inevitably lead to a series of notorious issues such as filter bubbles \cite{filter_bubble} and echo chamber \cite{echo_chamber}. Thus, it is crucial to consider the dynamic user-system interactions to alleviate Matthew effect. \emph{2) Preference Learning}. Prior studies \cite{MatthewEffect1_1,MatthewEffect1_2,MatthewEffect1_3,MatthewEffect1_4} show that the key to mitigating Matthew effect is to learn diverse user preferences. Thus, many methods leverage multiplex external Knowledge Graphs (KGs) to model multi-aspect preferences. But traditional KG edges are limited to linking only two vertices (\emph{i.e.}, factors), restricting preference learning to pairwise interactions. Instead, user relations exhibit intricate complexity, such as a user's preference for a garment involves multiple factors like color, brand, style, and texture simultaneously. Hence, extending the number of vertices for learning diverse preferences is rather important.\\
\indent To address these issues, we propose a novel paradigm, \textbf{Hy}pergraph-Enhanced Multi-Preference Learning for Alleviating Matthew Effect in \textbf{Co}nversational \textbf{Rec}ommendation (\textbf{HyCoRec}), which consists of Hypergraph-Enhanced Multi-Preference Learning and Hypergraph-aware CRS. The former aims to model multi-aspect preferences, specifically targeting item-aspect, entity-aspect, word-aspect, review-aspect, and knowledge-aspect preferences. It addresses the Matthew effect in CRS by utilizing item-based hypergraph, entity-based hypergraph, word-based hypergraph, item reviews, and knowledge graphs to learn and derive these preferences. The latter focuses on leveraging these multi-aspect preferences as users interact with the system. Concretely, multi-aspect preferences are adopting to accurately predict the next utterances in the conversational task, and effectively make diverse item predictions in the recommendation task. By incorporating and utilizing these multi-aspect preferences, the system aims to provide precise and diverse recommendations that cater to the individual user's preferences and needs for alleviating the Matthew effect as they continue to engage with the system. Empirically, extensive experimental results on two benchmarks show that HyCoRec outperforms all the compared baselines, and the superior of mitigating Matthew effect. \\
\indent Overall, our main contributions are included:
\vspace{-10pt}
\begin{itemize}[leftmargin=8pt]
\item To the best of our knowledge, this is the first work to model multi-aspect user preferences, \emph{i.e.}, item-, entity-, word-, review-, knowledge-aspect preference, to alleviate Matthew effect in the CRS.
\item We proposed a novel end-to-end framework, HyCoRec, which adopts the multi-aspect preferences to effectively generate responses in the conversational task and accurately predict itms in the recommendation task. 
\item Quantitative and qualitative experimental results on two CRS-based datasets exhibit superior performance of HyCoRec and the effectiveness of mitigating Matthew effect in the CRS.
\end{itemize}

\section{Related Work}
\subsection{Conversational Recommender System}
Unlike traditional recomemnder system \cite{TRS_1, TRS_new, TRS_2, IRS_1, IRS_2}, Conversational Recommender System \cite{CRS_1, CRS_2, CRS_3, CRS_4, CRS_5, CRS_6, CRS_7, CRS_8} aims to capture user preferences through dialogues and provide high-quality recommendations. Previous research on CRS can be broadly categorized into two main types: attribute-based CRS \cite{deng2021unified,lei2020estimation,lei2020interactive,ren2021learning,xu2021adapting} and generation-based CRS \cite{chen2019towards,deng2023unified,li2022user,zhou2020improving,C2CRS,shang2023multi}. Attribute-based CRS involves capturing user preferences by asking questions about item attributes and generating responses using pre-defined templates \cite{lei2020estimation}. But this strategy often neglects the importance of generating responses that resemble natural human language, which can negatively impact the user experience. On the other hand, generation-based CRS tackles this issue by utilizing the Seq2Seq architecture \cite{vaswani2017attention} to integrate both conversation and recommendation tasks to produce fluent and coherent human-like responses. Despite their effectiveness, they fail to model users' diverse preferences since the user-item interactions data is rather sparse and limited. In contrast, our work aims to model multi-aspect preferences for exploring user diverse intricate relation patterns.

\subsection{Matthew Effect in Recommendation}
Matthew effect is a notorious issue in RSs. Recently, Liu et al. \cite{MatthewEffect1} have substantiated the occurrence of the Matthew effect in YouTube's recommendation system. Besides, Wang et al. \cite{MatthewEffect_QA} undertook a rigorous quantitative analysis, offering valuable insights into the quantitative characteristics of the Matthew effect in recommender systems based on collaborative filtering. To alleviate Matthew effect, one common method is to consider recommendation diversity strongly advocated by researchers \cite{MatthewEffect1_1, MatthewEffect1_2, MatthewEffect1_3, MatthewEffect1_4}, and another critical perspective is by removing popularity bias, a factor that has been identified as a catalyst for its amplification \cite{MatthewEffect2_1}. But these methods predominantly focus on investigating the Matthew effect in the static recommendation settings without considering the user-system feedback loop. Instead, our HyCoRec aims to alleviate Matthew effect considering dynamic user-system feedback loop.

\section{HyCoRec}
\begin{figure*}[t]
    \centering
    \includegraphics[width=1\textwidth]{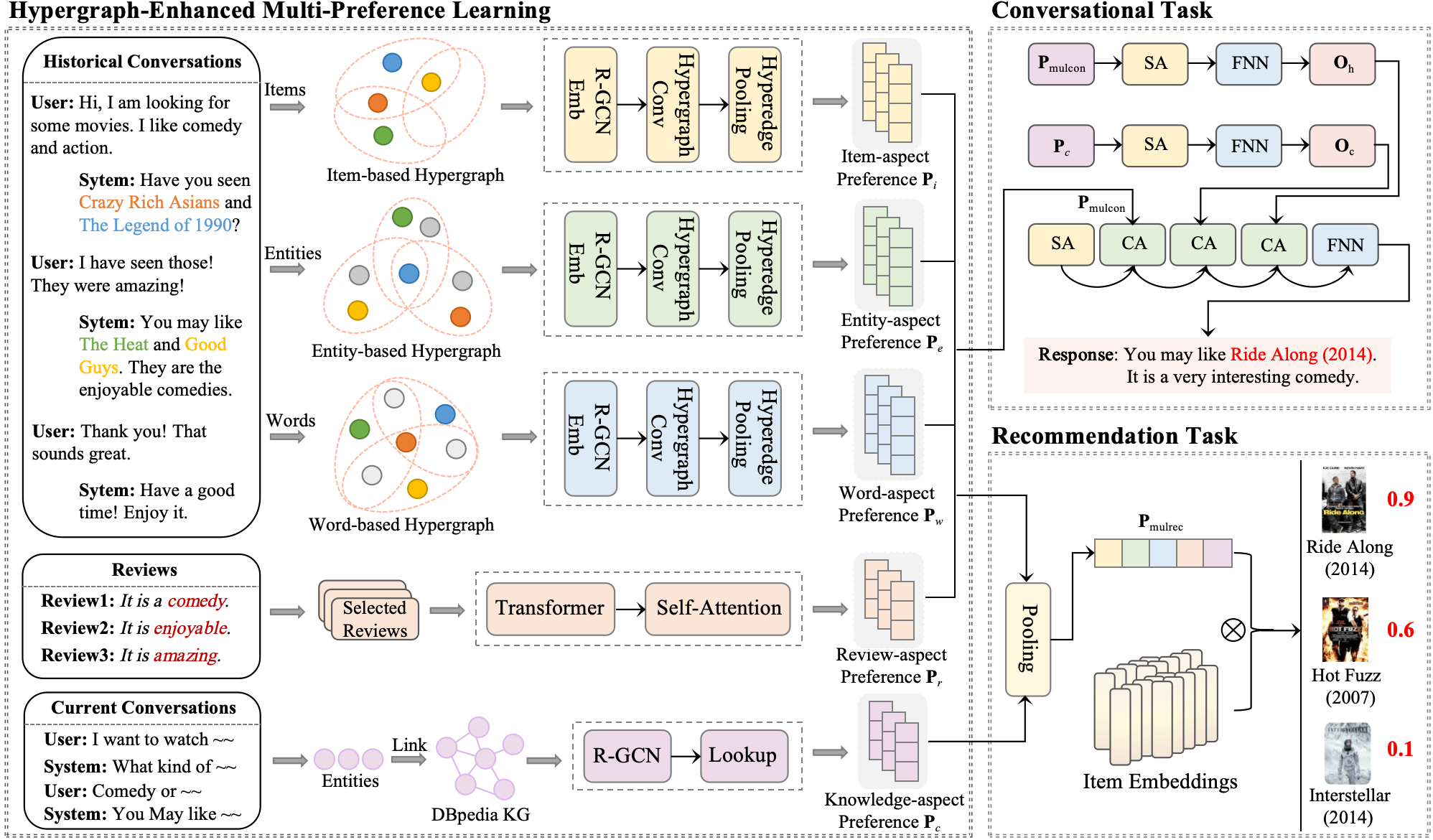} %HypergraphCRS4
    \caption{Overview of our HyCoRec framework, which consists of Hypergraph-Enhanced Multi-Preference Learning and Hypergraph-Aware CRS. The former aims to dynamically learn multi-aspect user preferences, while the latter contains the conversation task to generate diverse responses and the recommendation task to predict target items.}
    \label{fig:framework}
\end{figure*}

Matthew effect is a notorious issue in the CRS, and it inevitably becomes intensified over time due to the existence of the dynamic user-system feedback loop. To tackle these challenges, we propose a novel paradigm, HyCoRec, which consists of Hypergraph-Enhanced Multi-Preference Learning and Hypergraph-Aware CRS. The overall pipeline of our HyCoRec is depicted in Fig.\ref{fig:framework}.
\subsection{Preliminaries}
\subsubsection{Conversational Recommendation}
Conversational recommendation is a personalized approach where the system engages in continuous dialogues with users to gain a deeper understanding of their preferences and deliver customized suggestions. This interactive method enables the system to collect additional insights into the user's preferences, context, and requirements, resulting in more precise and relevant recommendations. 
CRSs are widely applied across diverse domains like e-commerce, music streaming, movie recommendations, and others, aiming to enrich user experience and satisfaction.

\subsubsection{Hypergraph}
Hypergraphs demonstrate intricate configurations, capturing intricate relationships among numerous elements through hyperlinks. In our research, we depict user inclinations by constructing multi-grained hypergraphs, encompassing the item-based hypergraph $\mathcal{G}^{(t)}_{\rm item}$, entity-based hypergraph $\mathcal{G}^{(t)}_{\rm entity}$, and the word-based hypergraph $\mathcal{G}^{(t)}_{\rm word}$. Each hypergraph can be delineated as $\mathcal{G}^{(t)}_{\rm item}=(\mathcal{I}^{(t)}_{*}, \mathcal{H}^{(t)}_{*}, \boldsymbol{\rm N}^{(t)}_{*})$, comprising: (1) a node collection $\mathcal{I}^{(t)}_{*}$; (2) a hyperege ensemble $\mathcal{H}^{(t)}_{*}$; (3) a $|\mathcal{I}^{(t)}_{*}| \times |\mathcal{H}^{(t)}_{*}|$ adjacent matrix $\boldsymbol{\rm N}^{(t)}_{*}$ signifying the weighted link between each node and hyperedge.

\subsection{Hypergraph-Enhanced Multi-Preference Learning}
Extensive experiments by most existing methods \cite{diverse_user_preference_1, diverse_user_preference_2, diverse_user_preference_3} have consistently shown that users with restricted preferences are highly influenced by the Matthew effect. Thus, the key to alleviating such bad effect is to model the diverse user preferences. Along this line, we formulate the Hypergraph-Enhanced Multi-Preference Learning, including Multi-Hypergraph Construction and Multi-Preference Learning.

\subsubsection{Multi-Hypergraph Construction}\label{sec:hypergraph}
Traditional KGs focus on pairwise interactions for preference learning, as edges connect only two vertices. However, user preferences often exhibit complex item relation patterns. To address this, we construct multiple hypergraphs (item-aspect, entity-aspect, and word-aspect), enabling connections between more than two vertices. 

\indent \textbf{Item-based Hypergraph.} Items directly reflect users' genuine preferences. Users might prefer related items, such as products from the same brand or with similar features. Thus, establishing connections among similar or functionally similar items is crucial for exploring a diverse preferences. To do this, we first extract items from a session and treat them as vertices, forming a hyperedge. Then, All hyperedges associated with a user are connected through shared items to create the item-based hypergraph $\mathcal{G}^{(t)}_{\rm item}$ as:
\begin{equation}
\begin{aligned}
\mathcal{G}^{(t)}_{\rm item}&=(\mathcal{I}^{(t)}_{i},\mathcal{H}^{(t)}_{i}, \boldsymbol{\rm N}^{(t)}_{i}).
\end{aligned}
\end{equation}where $\mathcal{I}^{(t)}_{i}$ means the item set extracted from the historical conversations, $\mathcal{H}^{(t)}_{i}$ is the hyperedge set, and $\boldsymbol{\rm N}^{(t)}_{i} \in \{0, 1\}^{|\mathcal{I}^{(t)}_{i}| \times |\mathcal{H}^{(t)}_{i}|}$ is the incidence matrix, which can be defined as:
\begin{equation}
\boldsymbol{\rm N}^{(t)}_{v,h} = \left\{
\begin{aligned}
& 1, && {\rm if} \quad v \in h \\
& 0, && {\rm if} \quad v \notin h\\
\end{aligned}
\right.
\label{incidence_matrix}
\end{equation}The degree of a vertex $v \in \mathcal{I}^{(t)}_{i}$ is denoted as $d(v) = \sum_{h \in \mathcal{H}^{(t)}_{i}}\boldsymbol{\rm N}^{(t)}_{v,h}$. Similarly, the degree of an edge $h \in \mathcal{H}^{(t)}_{i}$ is written as $\delta(h) = \sum_{v \in \mathcal{I}^{(t)}_{i}}\boldsymbol{\rm N}^{(t)}_{v,h}$. Besides, we use $\boldsymbol{\rm V}^{(t)}_{i}\in \mathbb{N}^{|\mathcal{I}^{(t)}_{i}| \times |\mathcal{I}^{(t)}_{i}|}$ and $\boldsymbol{\rm E}^{(t)}_{i} \in \mathbb{N}^{|\mathcal{H}^{(t)}_{i}| \times |\mathcal{H}^{(t)}_{i}|}$ to be the diagonal matrices of the vertex degrees and edge degrees, respectively.

\indent \textbf{Entity-based Hypergraph.} To address the sparsity and limitations of historical user-item interaction data, we utilize the extensive DBpedia KG \cite{DBpedia} to construct the entity-based hypergraph. Specifically, we extract individual items mentioned in conversations as entities and their $k$-hop neighbors to form each hyperedge. This approach allows us to capture shared semantic connotations among the extended neighbors. The hyperedges are then connected based on the entities they have in common. Formally, the entity-based hypergraph $\mathcal{G}^{(t)}_{\rm entity}$ can be represented as:
\begin{equation}
\begin{aligned}
\mathcal{G}^{(t)}_{\rm entity}&=(\mathcal{I}^{(t)}_{i} \cup \mathcal{I}^{(t)}_{e},\mathcal{H}^{(t)}_{e}, \boldsymbol{\rm N}^{(t)}_{e}).
\end{aligned}
\end{equation}where $\mathcal{I}^{(t)}_{e}$ represents the $k$-hop neighbors, $\mathcal{H}^{(t)}_{e}$ is the hyperedge set, and $\boldsymbol{\rm N}^{(t)}_{e} \in \{0, 1\}^{|\mathcal{I}^{(t)}_{e}| \times |\mathcal{H}^{(t)}_{e}|}$ is the incidence matrix defined by Eq.(\ref{incidence_matrix}). Similarly, $\boldsymbol{\rm V}^{(t)}_{e}\in \mathbb{N}^{|\mathcal{I}^{(t)}_{e}| \times |\mathcal{I}^{(t)}_{e}|}$ and $\boldsymbol{\rm E}^{(t)}_{e} \in \mathbb{N}^{|\mathcal{H}^{(t)}_{e}| \times |\mathcal{H}^{(t)}_{e}|}$ denote the diagonal matrices of vertex degrees and edge degrees, respectively.

\indent \textbf{Word-based Hypergraph.} Keywords in conversations are vital for understanding users' needs. By analyzing prominent words, we can identify specific preferences, which is crucial for modeling diverse user preferences. To achieve this, we build a word-based hypergraph using the word-oriented KG ConcetNet \cite{ConceptNet} to uncover semantic relations like synonymy, antonyms, and co-occurrence. We represent each historical conversation item as a keyword and extend it to include $k$-hop neighbors, forming a hyperedge. All the hyperedges connect through shared words. The word-based hypergraph $\mathcal{G}^{(t)}_{\rm word}$ can be defined as follows:
\begin{equation}
\begin{aligned}
\mathcal{G}^{(t)}_{\rm word}&=(\mathcal{I}^{(t)}_{i} \cup \mathcal{I}^{(t)}_{w}, \mathcal{H}^{(t)}_{w}, \boldsymbol{\rm N}^{(t)}_{w}).
\end{aligned}
\end{equation}where $\mathcal{I}^{(t)}_{w}$ is $k$-hop neighbors, $\mathcal{H}^{(t)}_{w}$ means the hyperedge set, and $\boldsymbol{\rm N}^{(t)}_{w} \in \{0, 1\}^{|\mathcal{I}^{(t)}_{w}| \times |\mathcal{H}^{(t)}_{w}|}$ is the incidence matrix defined as Eq.(\ref{incidence_matrix}). Similarly, let $\boldsymbol{\rm V}^{(t)}_{w}\in \mathbb{N}^{|\mathcal{I}^{(t)}_{w}| \times |\mathcal{I}^{(t)}_{w}|}$ and $\boldsymbol{\rm E}^{(t)}_{w} \in \mathbb{N}^{|\mathcal{H}^{(t)}_{w}| \times |\mathcal{H}^{(t)}_{w}|}$ denote the diagonal matrices of the vertex degrees and the edge degrees, respectively. 

\subsubsection{Multi-Preference Learning}
Upon constructing multiple hypergraphs as described earlier, we will leverage these hypergraphs to effectively capture diverse user preferences for mitigating the Matthew effect. This includes preferences related to items, entities, words, reviews, and knowledge aspects, all of which play a role in modeling multi-aspect preferences.

\indent \textbf{Item-aspect Preference.} 
Modeling item-aspect preferences holds significant importance in comprehending users' distinct tastes and preferences concerning the various items they interact with. In line with this objective, we derive the item-aspect preference $\boldsymbol{\rm P}_i$ by leveraging the item-based hypergraph. To effectively capture high-order relations, inspired by \cite{HyperCon}, we define our \emph{Hypergraph Convolution} function $\textsf{\rm HConv}(\cdot)$ as follows:
\begin{equation}
\begin{aligned}
\boldsymbol{\rm X}^{(l+1)}_m &=\textsf{\rm HConv} \left(\boldsymbol{\rm X}^{(l)}, \boldsymbol{\rm N}^{(t)}_{i}, \boldsymbol{\rm V}^{(t)}_{i}, \boldsymbol{\rm E}^{(t)}_{i}, \boldsymbol{\rm I}^{(t)}_{i}\right),\\
\textsf{\rm HConv}(\cdot) &= ({\boldsymbol{\rm V}^{(t)}_{i}})^{-1} \boldsymbol{\rm N}^{(t)}_{i} ({\boldsymbol{\rm E}^{(t)}_{i}})^{-1} ({\boldsymbol{\rm N}^{(t)}_{i}})^T \boldsymbol{\rm X}^{(l)} \boldsymbol{\rm W}^{(l)}_i,\\
\boldsymbol{\rm X}^{(l+1)} &= \textsf{\rm Pooling} \left(\boldsymbol{\rm X}^{(l+1)}_m \right)_{m=1}^{M}.
\end{aligned}
\end{equation}Here, $\boldsymbol{\rm X}^{l}$ and $\boldsymbol{\rm X}^{(l+1)}$ represent the input of the $l$-th and $(l+1)$-th layers, respectively, and $\boldsymbol{\rm W}^{(l)}_i$ denotes the trainable parameter. The notations $\boldsymbol{\rm N}^{(t)}_{i}$, $\boldsymbol{\rm V}^{(t)}_{i}$, and $\boldsymbol{\rm E}^{(t)}_{i}$ have been discussed in Section \ref{sec:hypergraph}. Specifically, $\boldsymbol{\rm I}^{(t)}_{i}$ signifies the item representations of $\mathcal{I}^{(t)}_{i}$ extracted from the encoded entity embeddings \cite{shang2023multi}. Additionally, $m$ denotes the number of heads in the multi-head architecture \cite{Trans}. Finally, we apply an average pooling $\textsf{\rm Pooling}(\cdot)$ on the representation $\boldsymbol{\rm X}^{(L+1)}$ obtained from the last layer (\emph{i.e.}, $(L+1)$ layer) to learn item-aspect preference $\boldsymbol{\rm P}_i$:
\begin{equation}
\begin{aligned}
\boldsymbol{\rm P}_i = \boldsymbol{\rm X}^{(L+1)} =\textsf{\rm Pooling} \left(\boldsymbol{\rm X}^{(L+1)}_m \right)_{m=1}^{M},
\end{aligned}
\end{equation}

\indent \textbf{Entity-aspect Preference.}
Capturing entity-aspect preferences is highly advantageous for unveiling complex relationship patterns underlying users' behaviors. To achieve this, we employ the entity-based hypergraph as a mechanism to learn entity-aspect preferences. In a similar vein to the item-aspect preference, the entity-aspect preference $\boldsymbol{\rm P}_e$ can be represented as:
\begin{equation}
\begin{aligned}
\boldsymbol{\rm X}^{(l+1)}_j &=\textsf{\rm HConv} \left(\boldsymbol{\rm X}^{(l)}, \boldsymbol{\rm N}^{(t)}_{e}, \boldsymbol{\rm V}^{(t)}_{e}, \boldsymbol{\rm E}^{(t)}_{e}, \boldsymbol{\rm I}^{(t)}_{i+e}\right),\\
\boldsymbol{\rm P}_e &=\boldsymbol{\rm X}^{(L+1)} = \textsf{\rm Pooling} \left(\boldsymbol{\rm X}^{(L+1)}_j \right)_{j=1}^{J}.
\end{aligned}
\end{equation}The specifics of $\boldsymbol{\rm N}^{(t)}_{e}$, $\boldsymbol{\rm V}^{(t)}_{e}$, and $\boldsymbol{\rm E}^{(t)}_{e}$ can be found in Section \ref{sec:hypergraph}. Besides, $\boldsymbol{\rm I}^{(t)}_{i+e}$ is the entity representations of the entity set $\mathcal{I}^{(t)}_{i} \cup \mathcal{I}^{(t)}_{e}$. Moreover, $j$ denotes the number of heads in the multi-head architecture, and $\boldsymbol{\rm W}^{(l)}_e$ is the trainable parameter. 

\indent \textbf{Word-aspect Preference.} Keywords occurring in conversations directly reflect users' specific or potential preferences. Derived from the word-based hypergraph, the word-aspect preference $\boldsymbol{\rm P}_w$ can be formulated as:

\begin{equation}
\begin{aligned}
\boldsymbol{\rm X}^{(l+1)}_f &=\textsf{\rm HConv} \left(\boldsymbol{\rm X}^{(l)}, \boldsymbol{\rm N}^{(t)}_{w}, \boldsymbol{\rm V}^{(t)}_{w}, \boldsymbol{\rm E}^{(t)}_{w}, \boldsymbol{\rm I}^{(t)}_{i+w}\right),\\
\boldsymbol{\rm P}_w &= \boldsymbol{\rm X}^{(L+1)} = \textsf{\rm Pooling} \left(\boldsymbol{\rm X}^{(L+1)}_f \right)_{f=1}^{F}.
\end{aligned}
\end{equation}Here $\boldsymbol{\rm N}^{(t)}_{w}$, $\boldsymbol{\rm V}^{(t)}_{w}$, and $\boldsymbol{\rm E}^{(t)}_{w}$ are explained in more detail in Section \ref{sec:hypergraph}. Additionally, $\boldsymbol{\rm I}^{(t)}_{i+w}$ represents the word representations obtained from the encoded entity embeddings. The variable $f$ denotes the number of heads in the multi-head architecture, and $\boldsymbol{\rm W}^{(l)}_f$ denotes the trainable parameter.

\indent \textbf{Review-aspect Preference.}
Item reviews provide valuable insights into users' experiences and reflections. Analyzing these reviews helps identify patterns, sentiment trends, and user attitudes, leading to a better understanding of user preferences. Taking inspiration from the merits of Transformer model, we utilize the Transformer framework to encode accessed reviews \cite{RevCore}. Specifically, given a review $R$, the output embeddings from the previous transformer layer, denoted as $\mathcal{T}^{l}(R)$, define the subsequent layer $\mathcal{T}^{l+1}(R)$ using the $\emph{Multi-head Attention}$ function $\textsf{MHA}(\cdot)$ as:
\begin{equation}
\begin{aligned}
&\mathcal{T}^{l+1}(R)=\textsf{MHA}(\mathcal{T}^{l}(R), \mathcal{T}^{l}(R), \mathcal{T}^{l}(R)),\\
&\textsf{MHA}(\boldsymbol{K}, \boldsymbol{Q}, \boldsymbol{V}) = [\textsf{head}^{l}_{1};\cdots;\textsf{head}^{l}_g]\boldsymbol{\rm W}^{l},\\
&\textsf{head}^{l}_{g}=\textsf{SA}(\mathcal{T}^{l}(R)\boldsymbol{\rm W}_k,\mathcal{T}^{l}(R)\boldsymbol{\rm W}_q, \mathcal{T}^{l}(R)\boldsymbol{\rm W}_v),\\
&\textsf{SA}(\boldsymbol{K},\boldsymbol{Q},\boldsymbol{V})=\textsf{Softmax}(\frac{\boldsymbol{Q}\boldsymbol{K}^{\rm T}}{\sqrt{d/g}})\boldsymbol{V},
\label{multi_head_attention}
\end{aligned}
\end{equation}where $g$ is the number of heads, $\boldsymbol{\rm W}^{l}$ denotes the trainable parameters, and each head $\textsf{head}^{l}_{g}$ is computed using the \emph{Scaled Dot-Product Attention} \cite{allneed} $\textsf{SA}(\cdot)$. $\boldsymbol{K}$, $\boldsymbol{Q}$ and $\boldsymbol{V}$ indicate the key, query and value matrices, respectively. $\boldsymbol{\rm W}_k$, $\boldsymbol{\rm W}_q$, and $\boldsymbol{\rm W}_v$ are learnable parameters. For convenience, we consider the output embeddings of the final transformer layer as the review-aspect preferences $\boldsymbol{\rm P}_r$:
\begin{equation}
\begin{aligned}
\boldsymbol{\rm P}_r=\textsf{MHA}(\mathcal{T}^{\mathcal{L}}(\mathcal{R}), \mathcal{T}^{\mathcal{L}}(\mathcal{R}), \mathcal{\mathcal{T}}^{L}(\mathcal{R})).
\label{multi_head_attention_1}
\end{aligned}
\end{equation}Here $\mathcal{L}$ is the number of transformer layers.

\indent \textbf{Knowledge-aspect Preference.} 
The information conveyed in the ongoing conversation reflects the dynamic preferences of the users, providing valuable insights into their current interests. Thus, our focus lies in modeling the preference for knowledge aspects by encoding the entities mentioned in the current conversation. Given the current conversation context $\mathcal{C}$, we leverage DBpedia and CN-DBpedia, to extract entities $\mathcal{E}_{k}=\{e_1, e_2, \cdots, e_k \}$ along the paths. To capture high-order entity representations, we use RGCN to explicitly capture relational semantics by adopt contrastive pre-training \cite{shang2023multi}. The representation of entity $e$ at the $(l+1)$-th layer can be computed as:
\begin{equation}
\boldsymbol{e}^{l+1} = \sigma (\sum_{r \in \mathcal{R}} \sum_{\hat{e} \in \mathcal{N}^r_{e}} \frac{1}{Z_l} \boldsymbol{\rm W}^l_1 {\hat{\boldsymbol{e}}}^{l} + \boldsymbol{\rm W}^l_2 \boldsymbol{e}^{l}),
\end{equation}where ${\boldsymbol{e}}^l$ is the $l$-th layer's representation of entity $e$, $\sigma$ means the sigmoid function, $\hat{e}$ refers to entities from the one-hop neighbor set $\mathcal{N}^r_{e}$ under relation $r$, and ${Z_l}$ is the hyperparameter. $\boldsymbol{\rm W}^l_1$ and $\boldsymbol{\rm W}^l_2$ can be trained. We use the representation $\boldsymbol{e}^L$ from the last layer as knowledge-aspect preference $\boldsymbol{\rm P}_c$:
\begin{equation}
\boldsymbol{\rm P}_c = {\rm RGCN}(\mathcal{E}_{k}) = \{\boldsymbol{e}^T_1, \boldsymbol{e}^T_2, \cdots, \boldsymbol{e}^T_k\},
\end{equation}where $\boldsymbol{e}^T_i$ is the embedding of $e_i$ via RGCN. 

\subsection{Hypergraph-Aware CRS} 
To combat the Matthew effect in the CRS, we adopt multi-aspect preferences, \emph{i.e.}, $\boldsymbol{\rm P}_i$, $\boldsymbol{\rm P}_e$, $\boldsymbol{\rm P}_w$, $\boldsymbol{\rm P}_r$, and $\boldsymbol{\rm P}_c$, to accurately predict item in the recommendation task and effectively generate responses in the conversational task.

\subsubsection{Recommendation Task}
The recommendation task aims to accurately predict items for users through natural conversations in dynamic user-system interactions. To address the Matthew effect, we first integrate multiple preferences to induce the fused preference $\boldsymbol{\rm P}_{\rm mulrec}$ in the recommendation task as:
\begin{equation}
\begin{aligned}
\boldsymbol{\rm P}_h &= [\boldsymbol{\rm P}_{i};\boldsymbol{\rm P}_{e};\boldsymbol{\rm P}_{w};\boldsymbol{\rm P}_{r}],\\
\boldsymbol{\rm P}_{\rm mulrec} &= \textsf{Pooling}([\textsf{Pooling}(\boldsymbol{\rm P}_{h});\boldsymbol{\rm P}_c]).
\end{aligned}
\label{diverse_p}
\end{equation}where $;$ denotes the concatenation operation. Next, the vector $\boldsymbol{\rm P}_{\rm mulrec}$ is used to select the suitable items in all the candidate set from item set $\mathcal{I}$, and the recommendation prediction is calculated as:
\begin{equation}
\begin{aligned}
\mathcal{P}_{\rm rec} = \textsf{Softmax}(\boldsymbol{\rm P}_{\rm mulrec} \cdot {\rm E}^T_{I}),
\end{aligned}
\end{equation}where ${\rm E}_{I}$ is embeddings of all candidate items from item set $\mathcal{I}$. We use cross-entropy loss \cite{shang2023multi} to learn the recommendation task:
\begin{equation}
\begin{aligned}
\mathcal{L}_{\rm r} = - \sum^{B}_{j=1} \sum^{|\mathcal{I}|}_{i=1} &[-(1-y_{ij}) \cdot {\rm log}(1-{\mathcal{P}}^{(j)}_{\rm rec}(i))\\
&+ y_{ij} \cdot {\rm log}({\mathcal{P}}^{(j)}_{\rm rec}(i))],
\end{aligned}
\end{equation}here the symbol $B$ represents the size of the mini-batch, and $y_{ij} \in \{0, 1\}$ denotes the target label.

\subsubsection{Conversational Task}
The conversation task focuses on generating proper dialogue utterances  to respond to user inputs. To generate diverse responses, we integrate multi-aspect preferences vectors to derive the fused preference in the conversation task $\boldsymbol{\rm P}_{\rm mulcon}$ as:
\begin{equation}
\begin{aligned}
\boldsymbol{\rm P}_{\rm mulcon} &= \textsf{MHA}([\boldsymbol{\rm P}_c; \boldsymbol{\rm P}_h; \boldsymbol{\rm P}_h]),
\label{multi_conv}
\end{aligned}
\end{equation}here $\boldsymbol{\rm P}_h$ is defined as Eq.(\ref{diverse_p}). Then, this fused preference $\boldsymbol{\rm P}_{\rm mulcon}$ is fed into the Transformer-based encoder-decoder framework for generating diverse responses. Let $\boldsymbol{\rm Y}^{n-1}$ be the output of the last time unit, then the current one $\boldsymbol{\rm Y}^{n}$ is:
\begin{equation}
\begin{aligned}
\boldsymbol{\rm A}^{n}_0 &= \textsf{MHA}(\boldsymbol{\rm Y}^{n-1}, \boldsymbol{\rm Y}^{n-1}, \boldsymbol{\rm Y}^{n-1}),\\
\boldsymbol{\rm A}^{n}_1 &= \textsf{MHA}(\boldsymbol{\rm A}^{n}_0, \boldsymbol{\rm P}_{\rm mulcon}, \boldsymbol{\rm P}_{\rm mulcon}),\\
\boldsymbol{\rm A}^{n}_2 &= \textsf{MHA}(\boldsymbol{\rm A}^{n}_1, \boldsymbol{\rm P}_c, \boldsymbol{\rm P}_c),\\
\boldsymbol{\rm A}^{n}_3 &= \textsf{MHA}(\boldsymbol{\rm A}^{n}_1, \boldsymbol{\rm P}_h, \boldsymbol{\rm P}_h),\\
\boldsymbol{\rm A}^{n}_4 &= \beta \cdot \boldsymbol{\rm A}^{n}_2 + (1 - \beta) \cdot \boldsymbol{\rm A}^{n}_3,\\
 \boldsymbol{\rm Y}^{n} &= \textsf{FFN}(\boldsymbol{\rm A}^{n}_4).
\label{response_generator}
\end{aligned}
\end{equation}Here $\textsf{FFN}(\cdot)$ is the fully-connected feed-forward network, and $\beta$ is hyper-parameter to balance two signals. To enhance the response diversity, we use preference-aware bias and item-related bias following \cite{shang2023multi}. Given the predicted sequence $\{s_{t-1}\}$, the probability of the next token is calculated as:
\begin{equation}
\begin{aligned}
\mathcal{P}_{\rm conv}( s_t | \{s_{t-1}\}) &= P_1(s_t|Y_i) + P_2(s_t|\boldsymbol{\rm P}_{\rm mulrec})\\
&+ P_3(s_t|\boldsymbol{\rm P}_{\rm mulrec}),\\
\label{conversation_loss_1}
\end{aligned}
\end{equation}where $s_t$ is the $t$-th utterances, and $\{s_{t-1}\}=s_1, s_2, \cdots, s_{t-1}$. Inspired by \cite{shang2023multi}, $P_1(\cdot)$, $P_2(\cdot)$, and $P_3(\cdot)$ are the vocabulary probability, vocabulary bias, and copy probability, respectively. Next, we use the cross-entropy loss:
\begin{equation}
\begin{aligned}
\mathcal{L}_{\rm cå} &= - \sum_{b=1}^B \sum_{t=1}^T {\rm log}(\mathcal{P}_{\rm conv}( s_t | \{s_{t-1}\})).
\label{conversation_loss_2}
\end{aligned}
\end{equation}Here $T$ denotes the truncated length of utterances. 

\begin{table*}
\small
\setlength{\tabcolsep}{8mm}
\setlength{\abovecaptionskip}{4pt}  
\centering
\renewcommand{\arraystretch}{1.0}
\begin{tabular}{l@{\hskip 0.00in}
l@{\hskip 0.13in}
c@{\hskip 0.08in}c@{\hskip 0.08in}c@{\hskip 0.08in}c@{\hskip 0.08in}c@{\hskip 0.08in}c@{\hskip 0in}
c@{\hskip 0.18in}
c@{\hskip 0.08in}c@{\hskip 0.08in}c@{\hskip 0.08in}c@{\hskip 0.08in}c@{\hskip 0.08in}c@{\hskip 0in}
c@{\hskip 0.13in}}
\toprule
\multirow{2}{*}{\textbf{}}&
\multirow{2}{*}{\textbf{Model}}&
\multicolumn{6}{c}{REDIAL}&
&
\multicolumn{6}{c}{TG-REDIAL}
&\\
\cline{3-8}
\cline{10-15}
\rule{0pt}{10pt}
&&R@10&R@50&M@10&M@50&N@10&N@50&&R@10&R@50&M@10&M@50&N@10&N@50\\
&TextCNN&0.0644&0.1821&0.0235&0.0285&0.0328&0.0580&
&0.0097&0.0208&0.0040&0.0045&0.0053&0.0077\\
&SASRec&0.1117&0.2329&0.0540&0.0593&0.0674&0.0936&
&0.0043&0.0178&0.0011&0.0017&0.0019&0.0047\\
&BERT4Rec&0.1285&0.3032&0.0475&0.0555&0.0663&0.1045&
&0.0043&0.0226&0.0013&0.0020&0.0020&0.0058\\
&ReDial&0.1705&0.3077&0.0677&0.0738&0.0925&0.1222&
&0.0038&0.0165&0.0012&0.0017&0.0018&0.0045\\
&TG-ReDial&0.1679&0.3327&0.0694&0.0771&0.0924&0.1286&
&0.0110&0.0174&0.0048&0.0050&0.0062&0.0076\\
&KBRD&0.1796&0.3421&0.0722&0.0800&0.0972&0.1333&
&0.0201&0.0501&0.0077&0.0090&0.0106&0.0171\\
&KGSF&0.1785&0.3690&0.0705&0.0796&0.0956&0.1379&
&0.0215&0.0643&0.0069&0.0087&0.0103&0.0194\\
&KGConvRec&0.1819&0.3587&0.0711&0.0794&0.0969&0.1358&
&0.0220&0.0524&0.0088&0.0102&0.0119&0.0185\\
&BERT&0.1608&0.3525&0.0597&0.0688&0.0831&0.1255&
&0.0040&0.0194&0.0011&0.0017&0.0018&0.0050\\
&XLNet&0.1569&0.3590&0.0583&0.0677&0.0811&0.1255&
&0.0040&0.0187&0.0011&0.0017&0.0017&0.0048\\
&BART&0.1693&0.3783&0.0646&0.0744&0.0888&0.1350&
&0.0047&0.0187&0.0012&0.0017&0.0020&0.0048\\
&MHIM&0.1966&0.3832&0.0742&0.0830&0.1027&0.1440&
&0.0300&0.0783&0.0108&0.0129&0.0152&0.0256\\
&\textbf{HyCoRec*}&\textbf{0.2231}&\textbf{0.4351}&\textbf{0.0797}&\textbf{0.0898}&\textbf{0.1123}&\textbf{0.1579}&
&\textbf{0.0377}&\textbf{0.0826}&\textbf{0.0154}&\textbf{0.0173}&\textbf{0.0162}&\textbf{0.0245}\\
\bottomrule
\end{tabular}
\caption{\label{tab:recommendation} Recommendation results. * indicates statistically significant improvement (\emph{p} < 0.05) over all baselines.}
\end{table*}

\section{Experiments and Analyses}
We conduct experiments to fully evaluate our HyCoRec and answer the following questions:
\vspace{-8pt}
\begin{itemize}
\item \textbf{RQ1:} How does HyCoRec perform compared with all baselines in the recommendation task?
\item \textbf{RQ2:} How does HyCoRec perform compared with all baselines in the conversation task?
\item \textbf{RQ3:} How does HyCoRec alleviate Matthew effect in the CRS?
\item \textbf{RQ4:} How do the item-based hypergraph $\mathcal{G}^{(t)}_{\rm item}$, entity-based hypergraph $\mathcal{G}^{(t)}_{\rm enti}$, word-based hypergraph $\mathcal{G}^{(t)}_{\rm word}$, and item reviews $R$ contribute to the performance?
\item \textbf{RQ5:} How do parameters affect our HyCoRec?
\item \textbf{RQ6:} It is better to provide the case studies to comprehensively understand about how HyCoRec handles Matthew effect in the CRS?
\end{itemize}

\subsection{Experimental Protocol}
\textbf{Datasets.} We evaluate our HyCoRec on two challenging CRS-based datasets REDIAL \cite{TDCR} and TG-REDIAL \cite{Topic-Guided}. The REDIAL consists of 11,348 dialogues involving 956 users and 6,924 items, while the TG-REDIAL contains 10,000 dialogues with 1,482 users and 33,834 items. The reviews in REDIAL are sourced from the IMDb\footnote{https://www.dbpedia.org/}, while the reviews in TG-REDIAL are collected from Douban\footnote{https://movie.douban.com/}.\\
\noindent \textbf{Baselines.} To fully evaluate our HyCoRec, we conduct a comprehensive evaluation by comparing our method with several state-of-the-art methods. The compared methods include \textbf{TextCNN} \cite{TextCNN}, \textbf{SASRec} \cite{SASRec}, \textbf{BERT4Rec} \cite{BERT4Rec}, \textbf{Transformer} \cite{Trans}, \textbf{ReDial} \cite{li2018towards}, \textbf{KBRD} \cite{chen2019towards}, \textbf{KGSF} \cite{zhou2020improving}, \textbf{KGConvRec} \cite{sarkar2020suggest}, \textbf{BERT} \cite{BERT}, \textbf{XLNet} \cite{XLNet}, \textbf{BART} \cite{BART}, \textbf{DialoGPT} \cite{DialoGPT}, \textbf{GPT-3} \cite{GPT3},  \textbf{C2-CRS} \cite{C2CRS},  \textbf{LOT-CRS} \cite{LOTCRS}, \textbf{UniCRS} \cite{UniCRS}, and \textbf{MHIM} \cite{shang2023multi}.

\subsection{Recommendation Performance (RQ1)}
Following \cite{shang2023multi}, we adopt Recall@\emph{K} (R@\emph{K}), MRR@\emph{K} (M@\emph{K}), NDCG@\emph{K} (N@\emph{K}) (\emph{K}=10, 50) to evaluate the recommendation task. Experimental results in Table \ref{tab:recommendation} validate that our HyCoRec outperforms all the compared methods. 

The improvement of HyCoRec over these baselines can be attributed to three reasons: (1) Incorporating external knowledge sources like DBpedia and ConceptNet into the CRS proves beneficial in exploring users' intricate behaviors, considering the sparse and limited nature of user-item interaction data. (2) Dialogues serve as a treasure trove of valuable information beyond the explicit user inputs. By considering the ongoing conversation between the user and the system, HyCoRec can capture the user's current context and understand their immediate needs. (3) Modeling multi-aspect preferences, including item-, entity-, word-, review-, and knowledge-aspect preferences, to enhance recommendation diversity and alleviate the Matthew effect as users interact with the system over time.

\subsection{Conversational Performance (RQ2)}
In the conversational task, we adopt Distinct \emph{n}-gram (Dist-\emph{n}) \cite{shang2023multi} (\emph{n}=2,3,4) to evaluate the diversity of generated responses. Table \ref{tab:conversation} summarizes the experimental results, it is observed that our HyCoRec is superior to all the compared baselines. We can observe that the performance rankings of the four baseline models remain consistent, with KBRD leading the way, followed by KGSF, Transformer, and finally ReDial. This can be attributed to the fact that KBRD leverages external knowledge sources to align the representations of items and words. Besides, KGSF enriches its decoder by incorporating cross-attention mechanisms along with embeddings from both entity- and word-level knowledge graphs (KGs). Nevertheless, Transformer and ReDial solely rely on token sequences, disregarding the user preferences that are concealed within the entities. 

Compared with these baselines, the improvement of HyCoRec can be attributed to the fact that: (1) HyCoRec considers multi-aspect user preferences to generate diverse responses that effectively align with the user's multi-level preferences for alleviating Matthew effect. (2) To accurately forecast the next utterance, we integrate the fused preference obtained from various aspects into a Transformer-based encoder-decoder framework to generate high-quality responses to meet users' dynamic needs and interests.

\begin{table}
\small
\setlength{\tabcolsep}{0.1mm}
\setlength{\abovecaptionskip}{2pt}  
\centering
\renewcommand{\arraystretch}{1.0}
\begin{tabular}{l@{\hskip 0.1in}
c@{\hskip 0.05in}c@{\hskip 0.05in}c@{\hskip 0in}
c@{\hskip 0.15in}
c@{\hskip 0.05in}c@{\hskip 0.05in}c@{\hskip 0in}
c@{\hskip 0.1in}}
\hline
\multirow{2}{*}{\textbf{Model}}&
\multicolumn{3}{c}{REDIAL}&
&
\multicolumn{3}{c}{TG-REDIAL}
&\\
\cline{2-4}
\cline{6-8}
\rule{0pt}{10pt}
&Dist-2&Dist-3&Dist-4&
&Dist-2&Dist-3&Dist-4&\\
\hline
ReDial&0.0214&0.0659&0.1333&
&0.2178&0.5136&0.7960&\\
Trans.&0.0538&0.1574&0.2696&
&0.2362&0.7063&1.1800&\\
KBRD&0.0765&0.3344&0.6100&
&0.8013&1.7840&2.5977&\\
KGSF&0.0572&0.2483&0.4349&
&0.3891&0.8868&1.3337&\\
C2-CRS&0.2623&0.3891&0.6202&
&0.5235&1.9961&2.9236&\\
UniCRS&0.2464&0.4273&0.5290&
&0.6252&2.2352&2.5194&\\
LOT-CRS&0.3312&0.6155&0.9248&
&0.9287&2.4880&3.4972&\\
DialoGPT&0.3542&0.6209&0.9482&
&1.1881&2.4269&3.9824&\\
GPT-3&0.3604&0.6399&0.9511&
&1.2255&2.5713&4.0713&\\
MHIM&0.3278&0.6204&\textbf{0.9629}&
&1.1100&2.3520&3.8200&\\
\textbf{HyCoRec*}&\textbf{0.3661}&\textbf{0.6434}&0.9523&
&\textbf{1.2590}&\textbf{2.6000}&\textbf{4.1210}\\
\hline
\end{tabular}
\caption{\label{tab:conversation} Conversation results. * indicates statistically significant improvement (\emph{p} < 0.05) over all baselines.}
\end{table}

\subsection{Study on Matthew Effect (RQ3)}
As our objective is to alleviate the Matthew effect in the CRS, we extensively examine the recommendation outcomes and compare them with the strongest baselines to assess whether HyCoRec can effectively mitigate Matthew effect. To mitigate the Matthew effect, the crucial factor is to enhance the diversity of the recommendation results. Thus, we adopt two commonly-used metrics \emph{Coverage}@k (C@k) and Isolation-Index (Iso-Index) to assess the extent of recommendation diversification by taking into account the distinctions among recommended items. The higher coverage value demonstrates its superior capability to encompass a larger portion of the recommendation space, encompassing items from various categories. A lower isolation-index value indicates a higher diversity in the recommended results. 

As shown in Table \ref{tab:Mattheweffect}, it is evident that our HyCoRec consistently achieves the highest values of \emph{Coverage} and the lowest isolation-index value across all datasets compared with the strongest baselines. For instance, on the REDIAL, our HyCoRec achieves substantial improvements of 102.72\%, 75.90\%, 144.35\%, and 63.75\% in terms of \emph{Cover@5} when compared to all the strong models, namely KBRD, KGSF, KGConvRec, and MHIM, respectively. The results show that HyCoRec effectively addresses isolation and ensures extensive coverage of recommended items by providing users with a broader choice range, validating the superiority in alleviating the Matthew effect as the user interacts with the system.

\begin{table}[tp]
    \small
	\setlength{\tabcolsep}{1.3mm}
	\setlength{\abovecaptionskip}{2pt}  
	\renewcommand{\arraystretch}{1.0}
	%\caption{Empirical comparisons under different settings for ablation study on both datasets. }
	\begin{center}
		\renewcommand{\arraystretch}{1.0}
		\begin{tabular}{>{\raggedright\arraybackslash}p{1.5cm}|ccccc}
             \hline
              {Datasets}&\multicolumn{5}{c}{REDIAL}\\
			\hline
		     Models&C@5&C@10&C@15&C@20&Iso-Index\\
			\hline
			KBRD & 0.0579& 0.0810&0.0961&0.1072&0.1149\\
			KGSF& 0.0664&0.0831&0.1195&0.1366&0.1055\\
              KGConvRec &0.0478&0.0735&0.1044&0.1235&0.1003\\
              MHIM &0.1098&0.1492&0.1747&0.1977&0.0923\\
             \textbf{HyCoRec}&\textbf{0.1168}&\textbf{0.1579}&\textbf{0.1848}&\textbf{0.2071}&\textbf{0.0617}\\
			\hline
	          {Datasets}&\multicolumn{5}{c}{TG-REDIAL}\\
			\hline
		     Models&C@5&C@10&C@15&C@20&Iso-Index\\
			\hline
			KBRD&0.0757& 0.1204&0.1468&0.1584&0.1222\\
			KGSF&0.0847&0.1324&0.1606&0.1858&0.1198\\
              KGConvRec &0.0720&0.0904&0.1228&0.1515&0.1091\\
              MHIM &0.1749&0.2493&0.2939&0.3423&0.1042\\
             \textbf{HyCoRec}&\textbf{0.1841}&\textbf{0.2743}&\textbf{0.3100}&\textbf{0.3608}&\textbf{0.0791}\\
			\hline
		\end{tabular}
	  \caption{\label{tab:Mattheweffect} Results on \emph{C@k} and \emph{Iso-Index} metrics.}
	\end{center}
\end{table}

\begin{table}
\small
\setlength{\tabcolsep}{0.1mm}
\setlength{\abovecaptionskip}{2pt}  
\centering
\begin{tabular}{l@{\hskip 0.1in}
c@{\hskip 0.05in}c@{\hskip 0in}
c@{\hskip 0.15in}
c@{\hskip 0.05in}c@{\hskip 0in}
c@{\hskip 0.1in}}
\toprule
\multirow{2}{*}{\textbf{Model}}&
\multicolumn{2}{c}{REDIAL}&
&
\multicolumn{2}{c}{TG-REDIAL}
&\\
\cline{2-3}
\cline{5-6}
\rule{0pt}{10pt}
&R@10&R@50&
&R@10&R@50&\\
\hline
\textbf{HyCoRec}&\textbf{0.2231}&\textbf{0.4351}&
&\textbf{0.0377}&\textbf{0.0826}&\\
\hline
w/o item hypergraph &0.2061&0.4177&
&0.0347&0.0733\\
w/o entity hypergraph&0.2049&0.4206&
&0.0267&0.0696\\
w/o word hypergraph&0.2058&0.4194&
&0.0257&0.0661\\
w/o item reviews&0.2102&0.4203&
&0.0267&0.0771\\
\bottomrule
\end{tabular}
\caption{\label{tab:ablation_studies} Ablation studies on the recommendation task.}
\end{table}

\subsection{Ablation Studies (RQ4)}
In this part, we conduct ablation experiments with different variants of HyCoRec to verify the contributions of each component, including: 1) w/o item hypergraph: we remove the item-based hypergraph; 2) w/o entity hypergraph: we remove the entity-based hypergraph; 3) w/o word hypergraph: we remove the word-based hypergraph; 4) w/o item reviews: we remove item reviews. As shown in Table \ref{tab:ablation_studies}, we can observe that a substantial decline in performance when removing any type of component. The main reason is that various knowledge data can effectively explore users' multi-aspect preferences. This observation highlights the effectiveness of HyCoRec in alleviating the Matthew effect by providing diverse recommendation results. 

\subsection{Hyperparameters Analysis (RQ5)}
\begin{figure}[t]
\centering
\subfigure{\label{fig:dimension_hr} \includegraphics[scale=0.28]{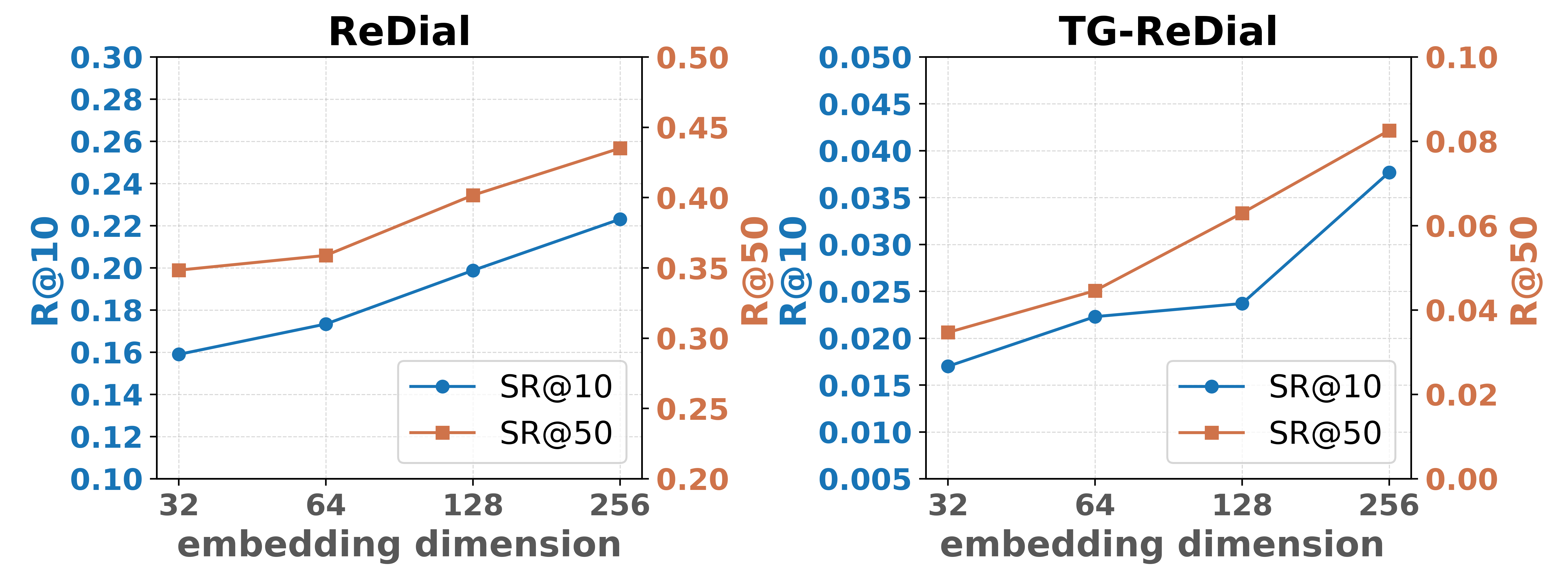}}
\subfigure{\label{fig:dimension_ndcg} \includegraphics[scale=0.28]{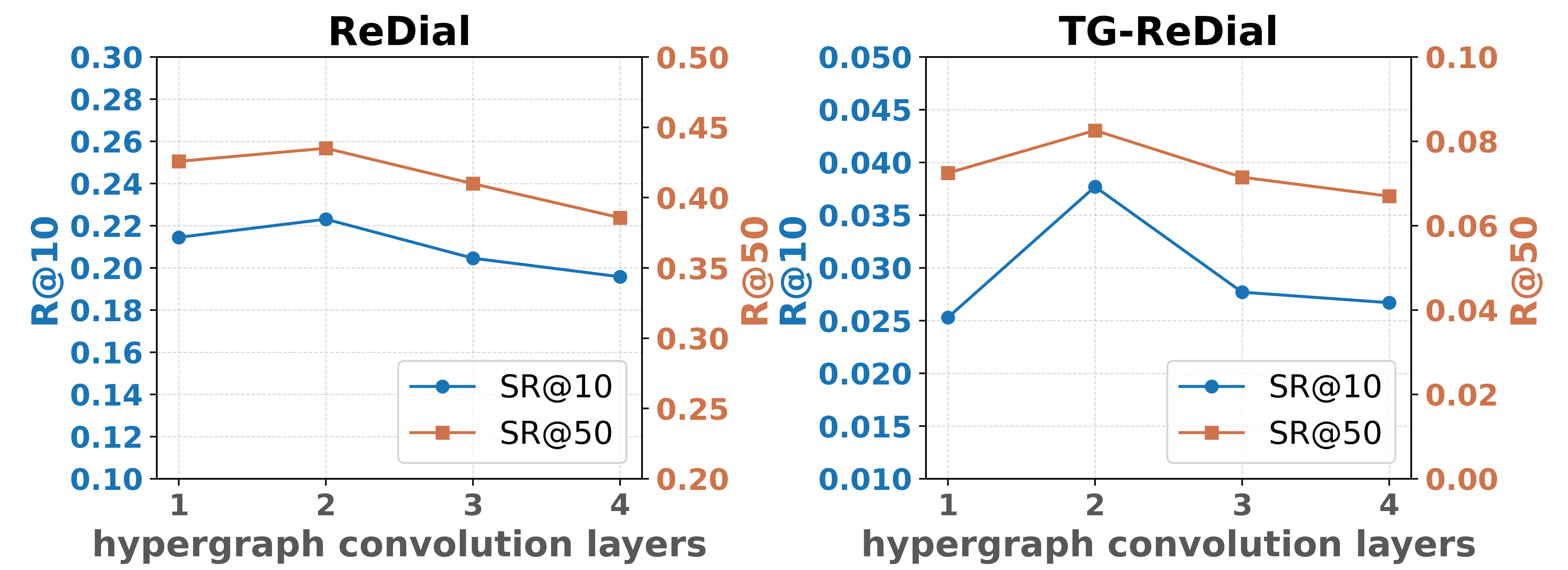}}
\subfigure{\label{fig:dimension_ndcg} \includegraphics[scale=0.28]{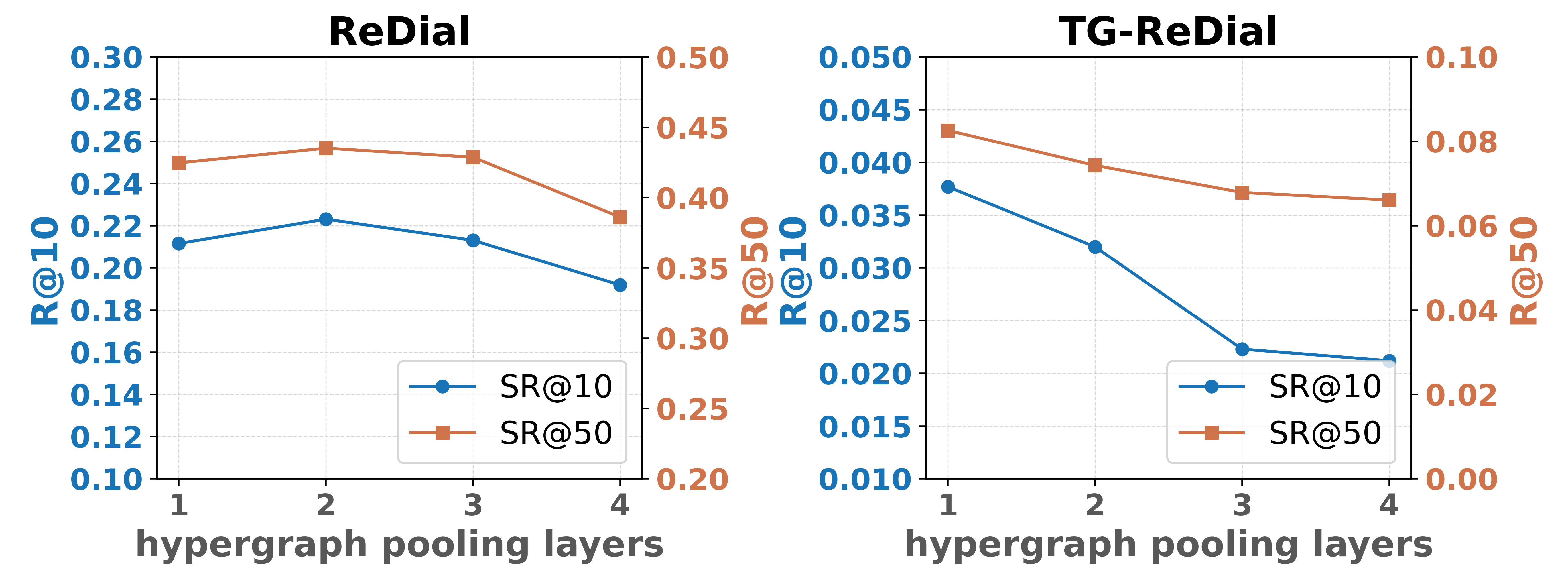}}
\caption{Impact of different hyperparameters.}
\label{fig:hyperparameters}
\end{figure}

Next, we investigate the impact of several important hyperparameters on the recommendation performance. As depicted in Fig.\ref{fig:hyperparameters}, we can observe:

Firstly, with the increase of embedding dimension, the recommendation performance continually improves. This is because the large dimension could encode sufficient high-level feature representations. Secondly, the model performance is optimal when the layer number is set to 2 on both datasets. The main reason is that larger hypergraph convolution layer numbers easily lead to model overfitting while the smaller one fail to capture enough feature representations. Lastly, a suitable hypergraph pooling layer number can enhance the model performance but the larger one might damage recommendation performance. The reason is that too large hypergraph pooling layer numbers might lose the important feature representations.

\subsection{Case Studies (RQ6)}
\begin{figure}[t]
    \centering
    \includegraphics[width=0.5\textwidth]{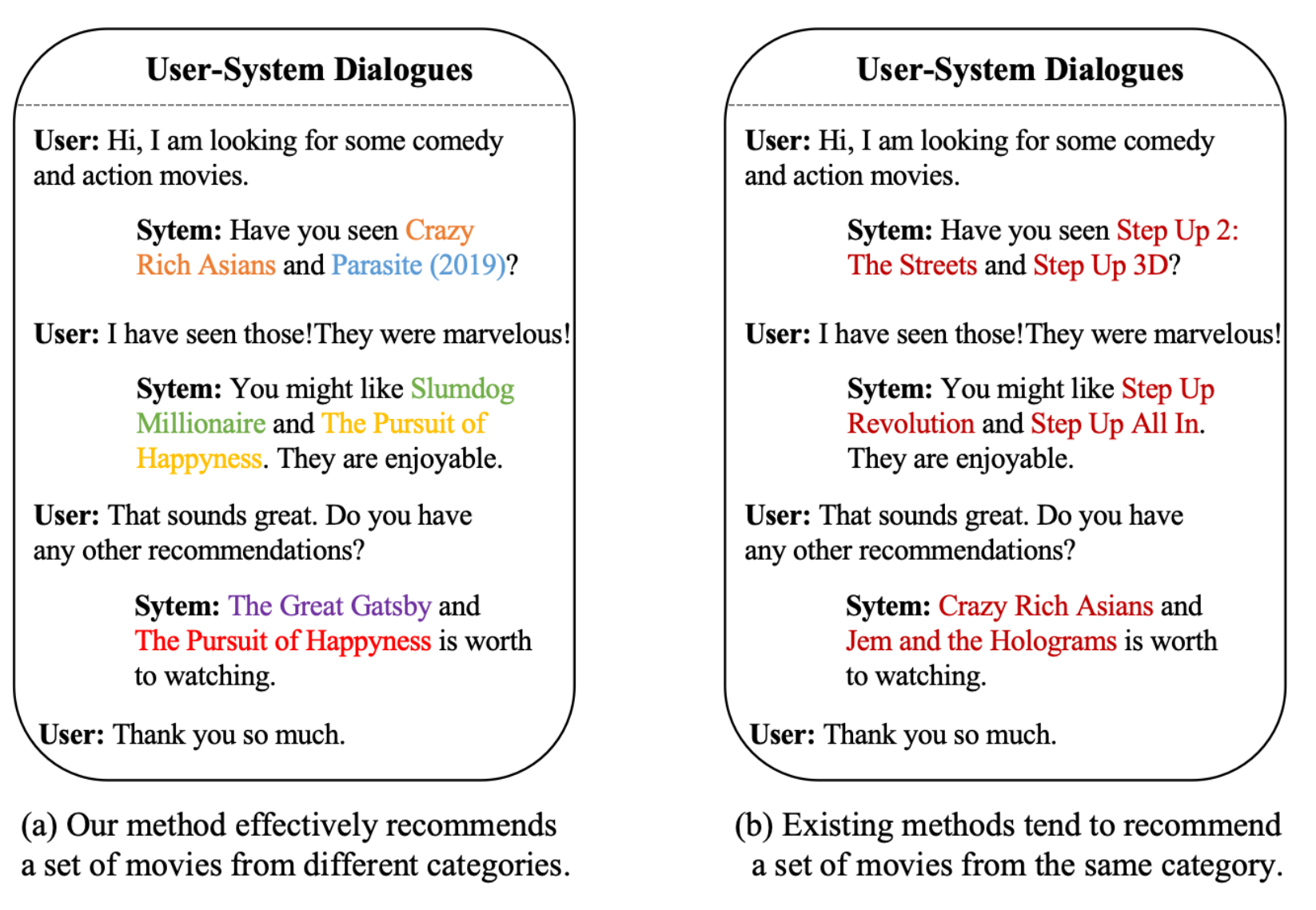} %HypergraphCRS4
    \caption{Case studies to comprehensively understand about how our proposed method HyCoRec handles Matthew effect in the CRS compared with most existing methods. Different colors denote different categories (see (a)) while the same color means the same category (see (b)).}
    \label{fig:case_studies}
\end{figure}

For a more in-depth understanding of how our proposed method, HyCoRec, tackles the Matthew effect during user-system interactions, we present comparative case studies between our approach and existing methods, visually illustrating the dialogue recommendation outcomes in human-computer interaction. As illustrated in Fig.\ref{fig:case_studies}, our method effectively recommends a diverse range of movies from different categories (see (a)), setting itself apart from existing methods that typically recommend movies from the same category (see (b)). The results highlight that our method achieves higher recommendation diversity, while most existing methods demonstrate lower diversity in recommendations. Generally, an effective strategy to alleviate the Matthew effect involves enhancing recommendation diversification \cite{MatthewEffect1_1,MatthewEffect1_2,MatthewEffect1_3,MatthewEffect1_4}. Thus, these results validate the effectiveness of our proposed method in mitigating the Matthew effect as users interact with the system over time in CRS. 

\section{Conclusion}
The Matthew effect is a notorious issue in the CRS, and it will be increasingly amplified due to the dynamic user-system feedback loop. To address these issues, we propose a novel paradigm, HyCoRec, which aims to learn multi-aspect user preferences, \emph{i.e.}, item-, entity-, word-, review-, and knowledge-aspect preferences, to effectively generate diverse responses in the conversation task and accurately predict items in the recommendation task for alleviating Matthew effect. Extensive experiments validate that our HyCoRec outperforms all the compared baselines and the superior of HyCoRec in alleviating Matthew effect in the CRS.

\section{Limitations}
While our HyCoRec has attained a remarkable state-of-the-art performance, it does have certain limitations. Firstly, the complexity and extensive nature of item reviews make the construction of the review-based hypergraph challenging and difficult. Consequently, the current version does not include the review-based hypergraph to capture a wider range of multiplex user relation patterns. Secondly, our proposed method necessitates the design of individual hypergraphs for learning multi-aspect preferences. This limitation could be addressed by developing a general framework that integrate any types of hypergraphs, thereby automatically unifying various knowledge sources.

\section{Ethics Statement}
The data utilized in our study are sourced from open-access repositories, and do not pose any privacy concerns. We are confident that our research adheres to the ethical standards set forth by ACL.

\section{Acknowledgements}
This work was supported in part by the National Key Research and Development Program of China under Grant
No.2021ZD0111601; National Natural Science Foundation of China under Grant No.62325605, Grant No.62206110 and Grant No.62206314; Guangzhou Basic Research Project for Basic and Applied Research under Grant No.202201010334; Guangdong Basic and Applied Basic Research Foundation under Grant No.2023A1515011374 and Grant No.2022A1515011835; Guangzhou Science and Technology Program under Grant No.2024A04J6365; Science and Technology Projects in Guangzhou under Grant No.2024A04J4388; and Guangdong Province Key Laboratory of Information Security Technology, Sun Yat-sen University. 

\bibliography{custom} 

\end{document}